\documentclass[epjCONF]{svjour}

\usepackage{graphics}
\usepackage[varg]{txfonts} 
\usepackage[latin1]{inputenc}
\session-title{A Universe of dwarf galaxies}

\begin{document}

\title{Ionized gas and sources of its ionization in the Irr galaxy IC 10}
\author{Egorov O. V.\inst{1}\fnmsep\thanks{\email{morzedon@gmail.com}} \and Arkhipova V. P.\inst{1}\fnmsep\thanks{\email{vera@sai.msu.ru}} \and Lozinskaya T. A.\inst{1}\fnmsep\thanks{\email{lozinsk@sai.msu.ru}}
 \and Moiseev A.
 V.\inst{2}\fnmsep\thanks{\email{moisav@gmail.com}}}

\institute{Sternberg Astronomical Institute, Moscow, Russia
 \and Special Astrophysical Observatory of Russian Academy of Science, Nizhnii Arkhyz, Karachai-Cherkessia, Russia}

\abstract{ IC 10 is the nearest starburst irregular galaxy
remarkable for its anomalously high number of WR stars. We report
the results of  an analysis   of the emission spectra of
HII-regions ionized by star clusters and  WR stars based on
observations made with the 6-m telescope of the Special
Astrophysical Observatory of the Russian Academy of Sciences using
MPFS field spectrograph and SCORPIO focal reducer operating in the
slit spectrograph mode.  We determine the masses and ages of
ionizing star clusters in the violent star-forming  region of the
galaxy in terms of the new evolutionary models of emission-line
spectra of HII-regions developed by \cite{RefMM}. We estimate the
amount of stars needed to ionize the gas in the brightest
HII-region HL 111 and report new determinations of oxygen
abundance in HII regions.}
%
\maketitle
\section{Introduction}
\label{intro} IC10 is the nearest dwarf Irr starburst galaxy; its H$_\alpha$ image
 appears as a giant complex of multiple shells and supershells with sizes ranging
  from 50 to 500-800 pc (see Fig.~1). About sixty star clusters have been found in
  this galaxy (\cite{RefHun}, \cite{RefShar}, \cite{RefTih}).
  The stellar population of the galaxy and its anomalously high space density of WR stars
   (similar to that in massive spiral galaxies) are indicative of a short recent starburst
    affecting the bulk of the galaxy.

\section{Observations}
\label{sec:1} We observed the galaxy with the 6-m telescope of
 the Special Astrophysical Observatory of the Russian Academy of Sciences using the MPFS
 field spectrograph and SCORPIO focal reducer operating in the Long Slit Spectrograph mode.
 Figure~1 shows the locations of MPFS fields (named according to the corresponding central
WR star)  and those of long-slit spectrograms (named according to
their position angle). Below we summarize the main results of
these observations, which we reported in detail in our papers
(\cite{RefArkh}, \cite{RefLoz}).

\section{Results}
\label{res}


We analyze the emission spectra of the ionized gas in the HII-regions observed, including
 the region of violent star formation (see Fig.1).  Figure~1 shows clusters from the lists
of \cite{RefHun},   \cite{RefTih} and \cite{RefShar} (circles indicate objects from
the former two lists and crosses,  those from the latter list). HII-regions
are labelled by their names according to the \cite{RefHL}
   catalog.

Earlier (\cite{RefLoz}) we found that diagnostic diagrams of the
relative line intensities from our observations with Long-Slit spectrograph
agree poorly with the
photoionization models available for the gas metallicity $Z=0.2Z\odot$ in IC10.

In this work we compare our observations with new evolutionary synthesis models of
\cite{RefMM} and find the diagnostic diagrams for these models to agree well with our
observations (see Fig.~2).

We use the evolutionary models of \cite{RefMM} to show in Fig.~3
the dependencies of the observed relative line intensities and
H$_\beta$-luminosity on cluster age for different cluster masses,
and also the dependence of ionization parameter on the
$[SII]6717/H_\alpha$ line intensity ratio for different ages and
masses. We use these dependencies to estimate the parameters of
the clusters that are the most likely sources of ionization for
the observed HII-regions.

\begin{table*}
\begin{center}
\caption{Parameters of HII-regions and their ionizing clusters.}
\label{tab:1}
\begin{tabular}{|c|c|c|c|c|c|c|}
\hline
HII-region & Cluster &  $N_{e}$, cm$^{-3}$ & L(H$\beta$), $10^{39}$ erg/s& Age, Myr  & Mass, $10^{5} M\odot$& $\log(U)$\\
    \hline
 HL111 & T54       & 70   & 1.3  & $ 3.5 \div 4.0 $& $1.0 \div 1.5 $ &-2.5 \\
 \hline
 HL106  & T50, T53  & 200  & 1.0  & $4.0 \div 4.5 $& $\ge 1.0$  & -3.0\\
 \hline
 HL50   & T32 & 30   & - & $3.4 \div 3.6$ or & $1.5 \div 2.0$ or& -2.47 \\
        & & &            & $2.5 \div 3.0$ & $0.4 \div 1.0$  & \\
 \hline
 HL111a,b, & T52   & 200  & - & $3.8 \div 4.5$ & $0.2 \div 0.6$ &
 -3.27 \\
 HL100 & & & & & & \\
 \hline
 HL46-48 & T24, T27&  -  &  -  & $4.0 \div 5.0 $& $0.2 \div 0.4$& -3.6\\
 \hline
 HL89  & T47 & 200 & - & $> 3$ & $\ge 0.2$ & -\\
 \hline
\end{tabular}
\end{center}
\end{table*}

\begin{table}
\caption{Oxygen abundances estimated with MPFS and long-slit spectrograph.}
\label{tab:1}
\begin{center}
\begin{tabular}{|c|c|}
\hline
HII-regions & $12+ log(O/H)$\\
\hline
\multicolumn{2}{|c|}{Observations with MPFS}\\
\hline
HL111c    &  $8.16 \pm 0.04$\\
HL111d   &  $8.17 \pm 0.04$\\
HL111e   &  $8.23 \pm 0.03$\\
\hline
HL100    &  $8.43 \pm 0.04$\\
 HL111a   &  $8.29 \pm 0.05$\\
\hline
HL106    &  $8.36 \pm 0.05$\\
 HL106a   &  $8.31 \pm 0.03$\\
\hline
HL97     &  $8.26 \pm 0.06$\\
 HL98     &  $8.35 \pm 0.05$\\
\hline
HL4      &  $8.30 \pm 0.16$\\
 HL6      &  $8.18 \pm 0.12$\\
\hline
\multicolumn{2}{|c|}{Observations with long-slit Spectrograph}\\
\hline
HL111a  & $8.33 \pm 0.04$\\
HL111b  & $8.41 \pm 0.01$ \\
\hline
HL111c  & $8.18 \pm 0.01$\\
HL111d  & $8.26 \pm 0.01$\\
HL111e  & $8.29 \pm 0.01$\\
\hline
HL106   & $8.27 \pm 0.05$\\
\hline
 SS     & $8.50 \pm 0.09$\\
\hline
HL37    & $8.15 \pm 0.12$\\
\hline
HL45    & $8.08 \pm 0.01$\\
\hline
HL50    & $8.18 \pm 0.08$\\
\hline
HL100   & $8.42 \pm 0.02$\\
\hline
HL89    & $8.44 \pm 0.04$\\
\hline
WR M10  & $8.42 \pm 0.02$\\
\hline
HL67    & $8.39 \pm 0.04$\\
\hline
HL41    & $8.38 \pm 0.02$\\
\hline
HL36    & $8.20 \pm 0.02$\\
\hline
HL22    & $8.38 \pm 0.01$\\
\hline
HL46-48 & $8.45 \pm 0.01$\\
\hline
\end{tabular}
\end{center}
\end{table}

Table~1 gives the masses, ages, and ionization parameters of the
clusters (named according to \cite{RefTih}) that are most
probable sources of ionization for the corresponding HII-regions.
The ages of ionizing clusters in IC~10 are shown to range from 2.5
to 5 Myr, and their masses, from $0.2\times10^{5}$ to $10^{5}
M\odot$.

We estimate the amount of O5V stars needed to ionize the gas in the HL111 nebula
based on the measured $H_\beta$-luminosity  and the photon UV-luminosity of O5V star
   $Q_0 =1.6\times10^{49}$ photons/s for the metallicity of
$Z=0.2Z\odot$ from \cite{RefSmith}.  We find that about a hundred of O5V stars are needed to
ionize HL111~-~the brightest HII-region
   in IC10.

We use the empirical method of \cite{RefPetPag} to determine the
oxygen abundance in the HII regions observed. The results are
presented in Table 2.

\begin{figure*}
\caption{Location of slit spectrograms and MPFS fields on the
H$_\alpha$-images of IC 10:  the entire galaxy (left) and the
bright region of current star formation (right). The asterisks denote the
spectroscopically confirmed WR stars from \cite{RefRoy} and from
\cite{RefMasHol}. The circles show the clusters from \cite{RefHun}
and from \cite{RefTih} and the crosses, the centers of clusters from
\cite{RefShar}. The names of HII-regions listed in Table~1 and
Table~2 are given according to \cite{RefHL}.} \label{fig:1}
\begin{center}\resizebox{1.8\columnwidth}{!}{\includegraphics{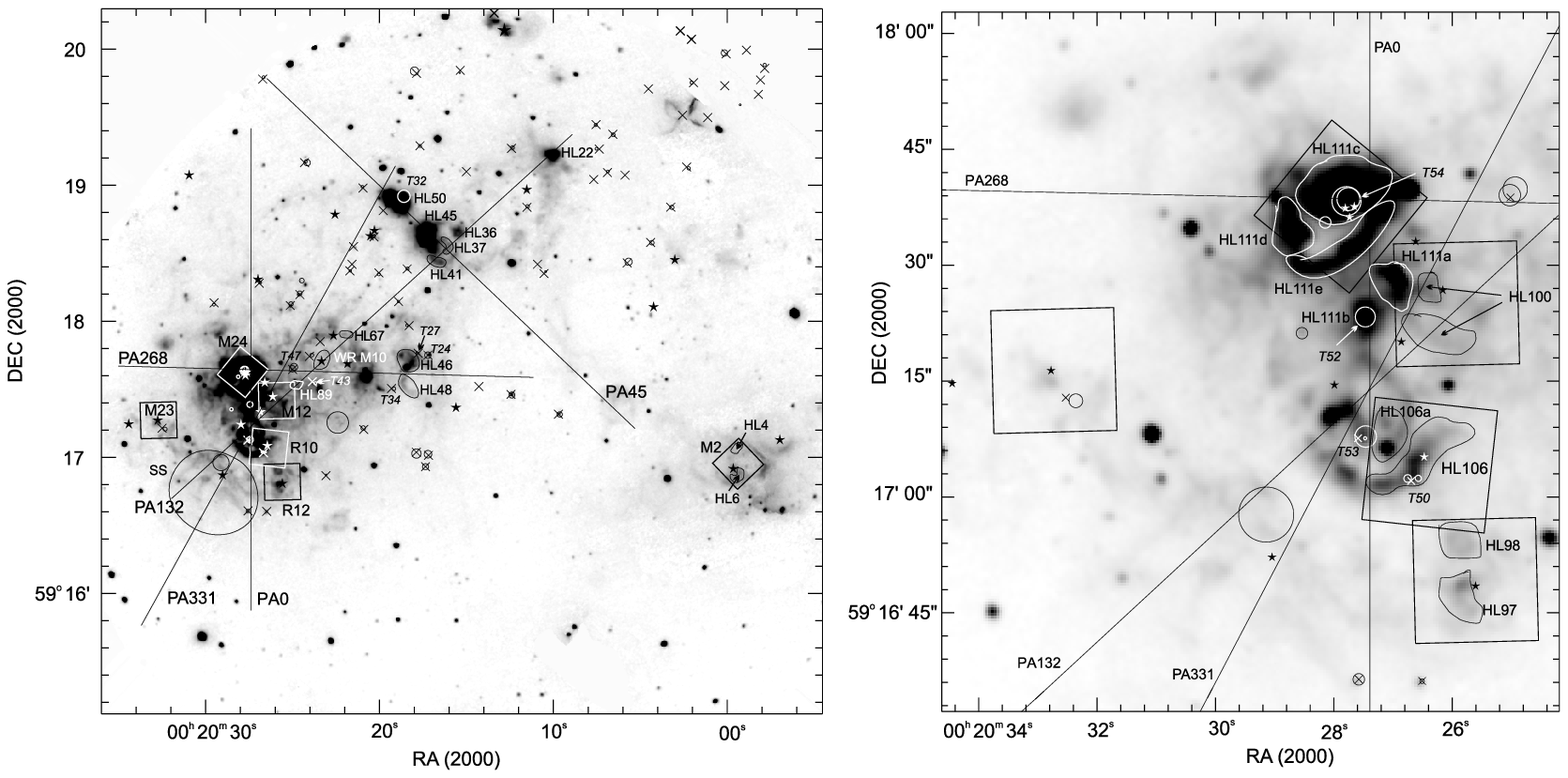}}
\end{center}
\end{figure*}

\begin{figure*}
\caption{Comparison of the results of our observations (open circles - Long Slit data,
filed circles - MPFS data) and those of \cite{RefMagr} (triangles) with
the diagnostic  diagrams
log([OIII]/H$_\beta$) vs log([NII]/H$_\alpha$) and log([SII]/H$_\alpha$) vs log([OIII]/H$_\beta$)
for models of \cite{RefMM} for z=0.004 and for different cluster masses and ages
(solid and dashed lines correspond to N$_e$ = 10 $cm^{-3}$ and N$_e$~=~100~$cm^{-3}$),
respectively.}
\label{fig:2}
\begin{center}
\resizebox{1.8\columnwidth}{!}{\includegraphics{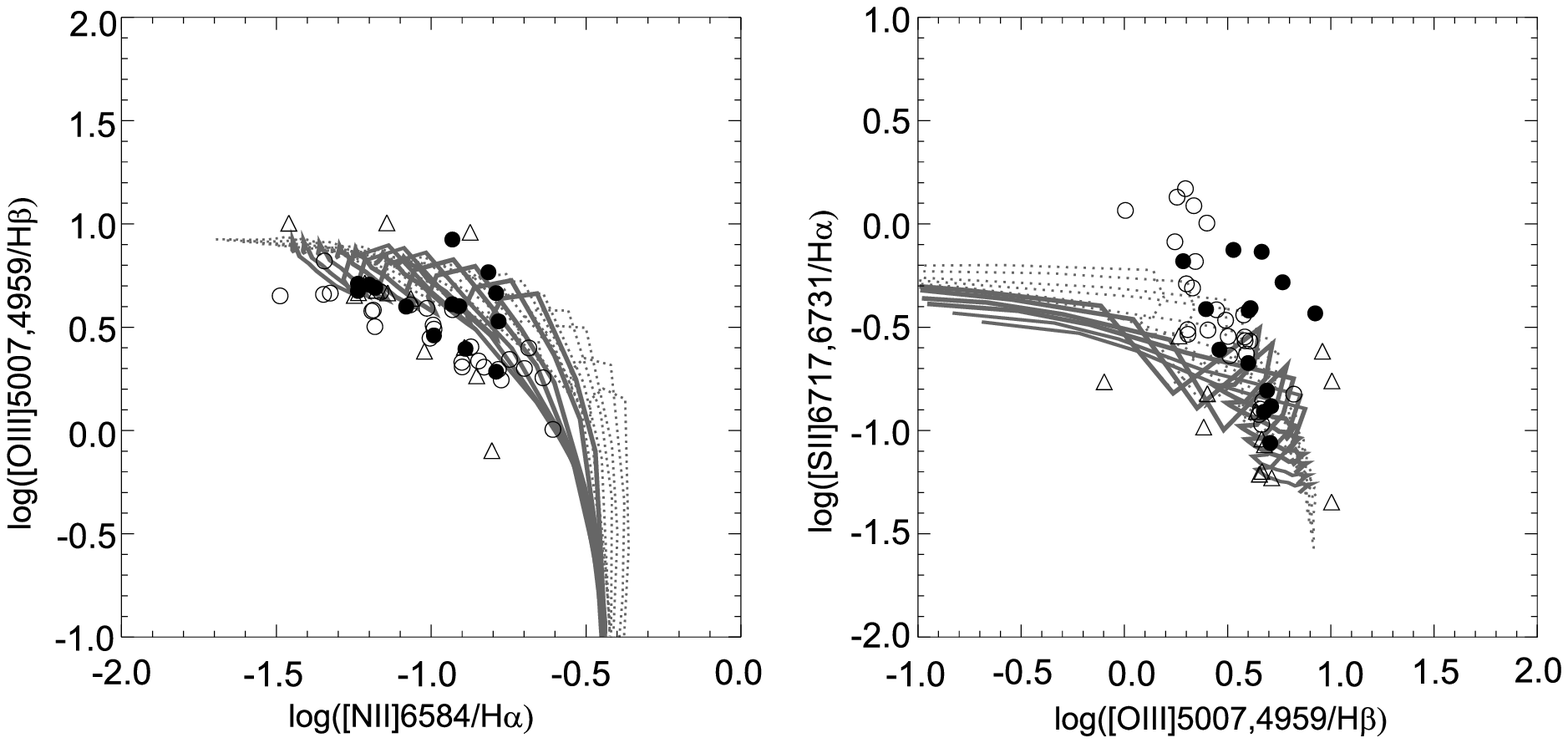}}
\end{center}
\end{figure*}

\section{Acknowledgments}
\label{acknowl}
This work was supported by the Russian Foundation of Basic Research
(project code 10-02-00091).
O.~V.~Egorov and A.~V.~Moiseev acknowledge the support from the Dynasty Foundation.
 The work was based on observations with the 6-m telescope of the Special Astrophysical
 Observatory, which is operated under the financial support from the Ministry of
Education and Science of the Russian Federation (registration number 01-43).

\begin{figure*}
\caption{Dependencies based on the data from Table 2b of the electronic version of
the paper \cite{RefMM} for N = 100 $cm^{-3}$ and metallicity z=0.004. Different curves
correspond to different cluster masses (indicated in the units of $10^{5} M\odot$).
The horizontal  and vertical dashed lines show our observational data for
the corresponding HII-regions.}
\label{fig:3}
\begin{center}
\resizebox{2.0\columnwidth}{!}{\includegraphics{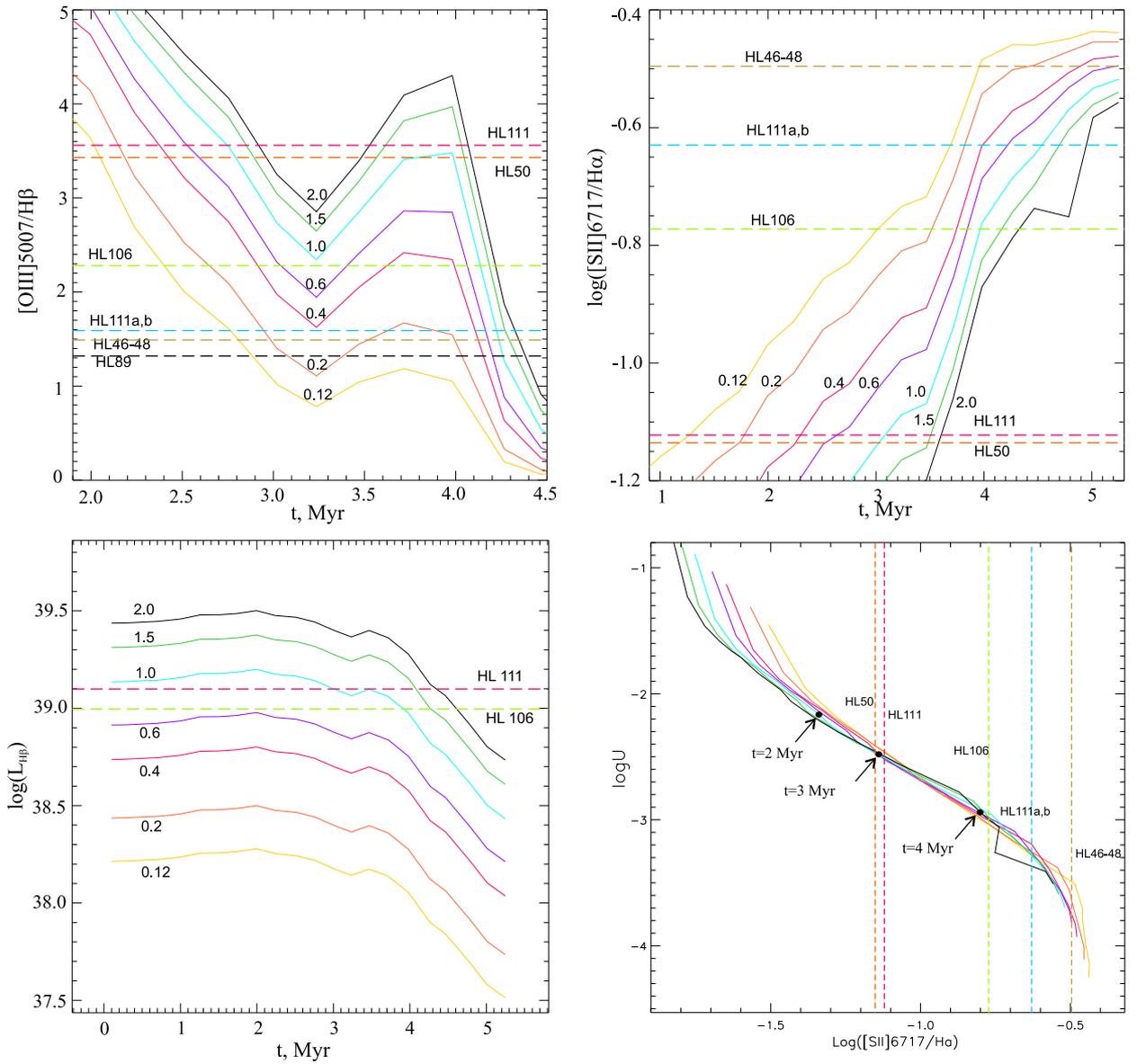}}
\end{center}
\end{figure*}


\begin{thebibliography}{}

\bibitem{RefArkh}
Arkhipova V. P., Egorov O. V., Lozinskaya T. A., Moiseev A. V., Astron. Lett. \textbf{36}, (2010), in press

\bibitem{RefHL}
Hodge P., Lee M. G., PASP \textbf{102}, (1990) 26

\bibitem{RefHun}
Hunter D. A., Astrophys. J. \textbf{559}, (2001) 225

\bibitem{RefLoz}
Lozinskaya T. A. et al., Astron. Lett. \textbf{35}, (2009) 811

\bibitem{RefMagr}
Magrini L. and Goncalves D. R., MNRAS \textbf{298}, (2009) 280

\bibitem{RefMM}
Martin-Manjon M. L. et al., MNRAS \textbf{403}, (2010) 2012-2032

\bibitem{RefMasHol}
Massey P. and Holmes S., Astrophys. J. Lett.\textbf{580}, (2002) L35

\bibitem{RefPetPag}
Pettini M. and Pagel B. E. J., MNRAS \textbf{348}, (2004) L59

\bibitem{RefRoy}
Royer P. et al., Astron. Astrophys. Lett. \textbf{366}, (2001) L1

\bibitem{RefShar}
Sharina M. E. et al., MNRAS \textbf{405}, (2010) 839-856

\bibitem{RefSmith}
Smith L. J. et al., MNRAS \textbf{337}, (2009) 1309-1328

\bibitem{RefTih}
Tikhonov N. A. and Galazutdinova O. A., Astro-Ph \textbf{1002.2046}, (2010)

\end{thebibliography}
\end{document}